\documentclass[onecolumn,12pt,preprintnumbers,amsmath,amssymb]{revtex4}
\usepackage{dcolumn}
\usepackage[dvips]{epsfig}
\usepackage{times}
\begin{document}       % INITIALIZE - DONT CHANGE % %  %

\date{\today}

\title{\bf Orbitally relieved magnetic frustration in NaVO$_{2}$}
\author{Ting Jia, Guoren Zhang and Zhi Zeng\footnote{Correspondence author: zzeng@theory.issp.ac.cn}}
%\email{zzeng@theory.issp.ac.cn}
\affiliation {Key Laboratory of Materials Physics, Institute of
Solid State Physics, Chinese Academy of Sciences, Hefei 230031, P.
R. China}
\author{H. Q. Lin}
\affiliation{\it Department of Physics and Institute of Theoretical
Physics, The Chinese University of Hong Kong, Shatin, Hong Kong,
P. R. China}
\date{\today}% It is always \today, today, but you may specify any date with \date.
\begin{abstract}
The magnetic properties of NaVO$_{2}$ are investigated
using full-potential linearized augmented plane wave method. We
perform calculations for three structures. For the rhombohedral
structure at 100 K, the \emph{t$_{2g}$}
orbitals of V ions are split into upper \emph{a$_{1g}$} and lower \emph{e'$_{g}$} orbitals
by a trigonal distortion of compression.
For the monoclinic structure at 91.5 K, the system behaves like a frustrated spin lattice
with spatially anisotropic exchange interactions.
For another monoclinic structure at 20 K, the magnetic frustration is relieved by a
lattice distortion which is driven by a certain orbital ordering,
and the long-range magnetic ordering is thus formed.
Moreover, the small magnetic moment originates from the
compensation of orbital moment for the spin moment.

\end{abstract}
\maketitle
\clearpage

\makeatletter
    \newcommand{\rmnum}[1]{\romannumeral #1}
    \newcommand{\Rmnum}[1]{\expandafter\@slowromancap\romannumeral #1@}
  \makeatother

\section{\bf Introduction}
The compounds with a common chemical formula \emph{AT}O$_{2}$ (\emph{A}=Na or Li, \emph{T}=3\emph{d}
transition metals) have been attracting a lot of attention for their large variety and richness in
physical phenomena\cite{Takada,Giot,Ezhov,Mazin,Darie,Reynaud}. Indeed, the discovery of supercondctivity
in the Na$_{0.35}$CoO$_{2}$$\cdot$1.3H$_{2}$O\cite{Takada}
and the application of LiCoO$_{2}$ in rechargeable Li batteries have accelerated investigations on their
fundamental physics. In addition, NaMnO$_{2}$ undergoes a structural
phase transition at 45 K to a long-range ordered antiferromagnetic (AFM) ground state\cite{Giot}, while NaNiO$_{2}$
exhibits ferromagnetic (FM) coupling in Ni-Ni plane below transition temperature\cite{Darie}.
Furthermore, a well-known member of this group, LiNiO$_{2}$, has no long-range magnetic ordering
even at low temperatures\cite{Reynaud,Chung,Reitsma,Mila}. The orbital frustration has been used to explain
the absence of magnetic ordering\cite{Reynaud,Reitsma,Mila},
and a local ordering of Ni$^{3+}$ Jahn-Teller (J-T)
orbitals is also proposed to be responsible for the complex magnetic properties\cite{Chung}.
Therefore, its controversial magnetic properties have been attracting considerable
interest.

These various phenomena always relate to the quasi-two-dimensional (2D)
triangular lattice formed by \emph{T} cations.
Such a triangular lattice may lead to magnetic frustration, since
all nearest-neighbor AFM interactions can not be satisfied simultaneously\cite{Giot,Ezhov,Mazin}.
Nevertheless, the magnetic frustration can be relieved by a certain orbital ordering (OO)\cite{Tokura,Yan,Horsch},
such as in LiVO$_{2}$\cite{Pen} and NaVO$_{2}$\cite{Mcqueen}.
 LiVO$_{2}$ has been found
to form a spin-singlet phase with corresponding OO at low temperatures\cite{Pen,Ezhov}.
Whereas, its sister compound NaVO$_{2}$ displays very different behaviours.

Recently, Onoda \emph{et al.}\cite{Onoda}
have revealed a superparamagnetic state driven by the short-range
ordered spin-1 (the total spin in one trimer S=1) trimerization\cite{Pen2} below the transition temperature
(\emph{T}=98 K). However,
 McQueen \emph{et al.} have reported two successive OO transitions in
NaVO$_{2}$\cite{Mcqueen}.
At 98 K, the system undergoes a continuous phase transition from
a rhombohedral (\emph{R}-3\emph{m}) phase
to a monoclinic (\emph{C}2/\emph{m}) one, corresponding
to the proposed OO of one electron per V$^{3+}$. Below 93 K, the system undergoes a discontinuous phase
transition to another monoclinic (\emph{C}2/\emph{m}) phase,
consistent with the proposed OO of two electron per V$^{3+}$.
In addition, a long-range ordered AFM state is formed at low temperatures,
while the magnetic moment observed in the ordered phase is about 0.98 $\mu$$_{B}$,
much smaller than the expected value (2 $\mu$$_{B}$).
The controversial magnetic states below 98 K obtained by these two groups bring us interests,
and the puzzling magnetic moment deserves to be explored.
There are no theoretical reports yet  to the best of our knowledge,
therefore we expect to understand its magnetic
properties at low temperatures upon our theoretical efforts.

In the present work, we have performed first-principles calculations to investigate
the electronic structures of NaVO$_{2}$, further to reveal the most possible
orbital and magnetic ordering, and to explore the origin of small magnetic moment
observed at low temperatures. Partially in agreement with the experimental
findings\cite{Mcqueen}, the OO, accompanied by a long-range magnetic ordering,
is found for the second monoclinic structure.
And the observed magnetic moment can be explained by including the spin-orbit coupling (SOC) interactions.

  This paper is organized as follows. The crystal structure and computational
details are described briefly in Sec. \Rmnum{2}. And the results and discussions are presented in Sec. \Rmnum{3}.
Finally, a brief conclusion is summarized in Sec. \Rmnum{4}.

\section{\bf Crystal Structure and Computational Details}
The lattice parameters  provided by McQueen \emph{et al.} are listed in Table \Rmnum{1}.
The structure of NaVO$_{2}$ is composed of 2D triangular-lattice
VO$_{2}$ layers of edge sharing VO$_{6}$ octahedra separated by sodium ions,
 which is rhombohedral (\emph{R}-3\emph{m}) at relatively high temperature (HT) (\emph{T}$>$98 K) and
monoclinic (\emph{C}2/\emph{m}) in both intermediate temperature (IT) (91.5 K$<$\emph{T}$<$98 K) and
low temperature (LT) (\emph{T}$<$91.5 K) phases\cite{Mcqueen}.
At 100 K,
 the V-O distances are nearly 2.04 {\AA}, but the O-V-O angle $\alpha$
is only \emph{85.53$^{\circ}$} (Fig. 1).
With lowering the temperature, $\alpha$ further reduces to \emph{85.42$^{\circ}$} at 91.5 K and
\emph{85.13$^{\circ}$} at 20 K. Therefore,
even in the HT phase, VO$_{6}$ octahedra in NaVO$_{2}$ have been different from the regular ones
under a trigonal distortion of compression along the threefold (111) axis\cite{Mcqueen}, which
induces the lowering of the \emph{O$_{h}$} local symmetry to \emph{D$_{3d}$}.
In VO$_{2}$ layers, the V-V geometry is built by two long (3.00781 {\AA})
and four short (2.99316 {\AA}) bonds in the IT phase,
 and reversely is by two short (2.97551 {\AA}) and four long (3.00526 {\AA}) ones
 in the LT phase\cite{interpretation}.
 The interlayer V-V distance is about 5.6 {\AA}.

All the calculations were performed by using the standard full-potential
linearized augmented plane wave code WIEN2k\cite{Blaha}.
The muffin-tin sphere radii of 2.22, 2.00 and 1.77 a.u. were chosen for the Na, V and O atoms, respectively.
The cutoff parameter
R$_{mt}$K$_{max}$ was chosen to be 7.0 and 100 \emph{k}-points were used over the first Brillouin zone.
The local spin density approximation (LSDA) of Perdew and Wang\cite{Perdew}
was used for the  exchange and correlation potential.
In order to take the strong-correlated nature of 3\emph{d} electrons into account explicitly,
we performed LSDA+U calculations\cite{Anisimov}, where
\emph{U$_{eff}$}=\emph{U}-\emph{J} (\emph{U} and \emph{J}
are on-site Coulomb and exchange interaction respectively) was used
instead of \emph{U}\cite{Dudarev}. And the orbital-dependent potential has the form of
$\Delta$\emph{V}$_{FLL}$ =
 -\emph{U$_{eff}$}(\emph{$\hat{n}$}$^{\sigma}$ - $\frac{1}{2}$\emph{I})\cite{Madsen},
 where \emph{$\hat{n}$}$^{\sigma}$ is the orbital occupation matrix of spin $\sigma$.
 This type of double-counting correction (DCC)
 has been called the fully localized limit\cite{Anisimov2,Laskowski}.
For NaVO$_{2}$, we used \emph{U$_{eff}$}=3.6 eV
 which has been used in its sister compound LiVO$_{2}$\cite{Ezhov}.
Note also that the conclusion made in this paper
is not affected for \emph{U$_{eff}$}=2-6 eV\cite{interpretation2}.
To explore the origin of small magnetic moment observed in the LT phase, we
performed LSDA+SOC+U calculations, where the SOC is included by the second-variational
method with scalar relativistic wave functions\cite{Blaha}.
The easy magnetization direction was set along (\={1}10) direction (short V-V bonds in the LT phase)
observed in the experiment\cite{Mcqueen}.

 In order to investigate different
 magnetic patterns, 2$\times$2$\times$2 supercell was used in our calculations. We took into account two
 AFM structures in V-V plane as described in Fig. 2(a): \Rmnum{1}-type antiferromagnetism (Fig. 2(a)(i))
is AFM exchange along the (010) and (\={1}10) directions with FM exchange along the (100) direction,
\Rmnum{2}-type antiferromagnetism (Fig. 2(a)(ii))
is AFM exchange along the (100) and (010) directions with FM exchange along the (\={1}10) direction.
Totally there were five possible magnetic configurations in our calculations for the IT and LT phases (Fig. 2(b)):
ferromagnetic (FM), \emph{C}-AF\Rmnum{1} (\Rmnum{1}-type antiferromagnetism
in plane, FM stacking), \emph{C}-AF\Rmnum{2} (\Rmnum{2}-type antiferromagnetism
in plane, FM stacking), \emph{G}-AF\Rmnum{1} (\Rmnum{1}-type antiferromagnetism
in plane, AFM stacking), \emph{G}-AF\Rmnum{2} (\Rmnum{2}-type antiferromagnetism
in plane, AFM stacking).

\section{\bf Results and discussions}

\subsection{\bf HT phase}
As the paramagnetic behavior
of NaVO$_{2}$ has been determined from the magnetic
 susceptibility measurements in the HT phase\cite{Mcqueen,Onoda},
 we just focus on the electronic structure instead of its magnetic properties.

The band structures obtained from LSDA and LSDA+U calculations are shown in Fig. 3.
 Within LSDA (Fig. 3(a)),
the bands near the Fermi level (\emph{E$_{F}$}) are mainly derived from V 3\emph{d} states.
 Since straight V-O-V paths are not present in layered NaVO$_{2}$ and instead only
nearly \emph{90$^{\circ}$} V-O-V bonds exit, the V 3\emph{d} states are quite narrow.
In the approximately octahedral crystal field,
the \emph{3d} orbitals are split into upper \emph{e$_{g}$} and lower
\emph{t$_{2g}$} states. As shown in Fig. 3(a),
 \emph{e$_{g}$} derived bands range from 1.5 to 2.5 eV and
\emph{t$_{2g}$} derived bands lie between -1.5 and 0.5 eV.
 The splitting between \emph{t$_{2g}$} and \emph{e$_{g}$} bands is about
1 eV. Under the trigonal crystal field, the \emph{t$_{2g}$}
orbitals are further split into one \emph{a$_{1g}$} and two degenerate \emph{e'$_{g}$}
orbitals. However, the splitting is much less than the band widths so that the \emph{t$_{2g}$}
orbitals still cross the \emph{E$_{F}$}, which denotes a metallic state within LSDA.
That is to say, LSDA calculations can not reproduce the insulating nature
of NaVO$_{2}$ from experiment\cite{Mcqueen}. The LSDA+U
 scheme\cite{Anisimov} is thus used to count the strong correlation of V 3\emph{d} electrons.
 As shown in Fig. 3(b), the empty \emph{a$_{1g}$} band is pushed upwards by about 1 eV, and a gap
 is opened near the \emph{E$_{F}$}. The system is hence an insulator due to electron correlation
 and NaVO$_{2}$ is indeed a good candidate for Mott-Hubbard insulator.

 According to the pure crystal field theory,
 the \emph{a$_{1g}$} orbital is of lower energy than the \emph{e'$_{g}$} orbitals under the
 trigonal distortion of  compression, which is opposite to our LSDA+U results.
So it is necessary to discuss the controversy on relative
 order of \emph{a$_{1g}$}-\emph{e'$_{g}$} in such trigonal distortions.
 In Ref. \cite{Landron}, Landron and Lepetit pointed out that this relative order
is strongly influenced by the \emph{e$_{g}$}-\emph{e'$_{g}$} hybridization.
The \emph{e$_{g}$} and \emph{e'$_{g}$} orbitals belong to the same irreducible representation (\emph{E$_{g}$})
and can thus mix despite the large \emph{t$_{2g}$}-\emph{e$_{g}$} energy difference.
Such a mix may be small but it modulates large energetic factors:
the on-site Coulomb repulsions. When the \emph{e$_{g}$}-\emph{e'$_{g}$} hybridization
is taken into account, the energy difference $\Delta$\emph{E} between the \emph{a$_{1g}$} and \emph{e'$_{g}$}
orbitals depends on two competitive parts: $\Delta$\emph{E} = $\Delta$\emph{E$_{1}$} + $\Delta$\emph{E$_{2}$} =
$\varepsilon$(\emph{a$_{1g}$}) - $\varepsilon$(\emph{e'$_{g}$}).
$\Delta$\emph{E$_{1}$} includes the kinetic energy,
the electron-charge interaction, and the interaction with the core electrons. $\Delta$\emph{E$_{2}$}
denotes the repulsion and exchange terms within the 3\emph{d} shells. Additionally,
$\Delta$\emph{E$_{1}$} and $\Delta$\emph{E$_{2}$} both depend on the amplitude of the
trigonal distortion and are of opposite effect with each other.
 Under a trigonal distortion of compression,
 if we only consider the crystal field effect ($\Delta$\emph{E$_{1}$}),
 the \emph{a$_{1g}$} orbital is of lower energy than the \emph{e'$_{g}$} orbitals ($\Delta$\emph{E} $<$ 0).
  Whereas if we take $\Delta$\emph{E$_{2}$} into account,
  the relative order between the \emph{a$_{1g}$} and \emph{e'$_{g}$}
  orbital is reversed ($\Delta$\emph{E} $>$ 0), comparing with the crystal field prediction.
Therefore, LSDA+U calculations predict that
the \emph{a$_{1g}$} orbital is of higher energy than the \emph{e'$_{g}$} orbitals in NaVO$_{2}$.
 In fact, such a controversy has been presented
 in another compressed triangular system Na$_{x}$CoO$_{2}$\cite{Landron2,Zou,Kashibae}. From
 the crystal field theory, some authors\cite{Kashibae} proposed that the energy of
 \emph{a$_{1g}$} orbital is lower than the \emph{e'$_{g}$} orbitals.
 However, the LDA+U method\cite{Landron2,Zou} yielded an
 \emph{a$_{1g}$} orbital of higher energy than the \emph{e'$_{g}$} orbitals.
Later, the experimental results\cite{Hasan} showed that the Fermi surface of the CoO$_{2}$ layers
 issues from the \emph{a$_{1g}$} orbital, not at all from the \emph{e'$_{g}$} orbitals,
 supporting the LDA+U results.

 \subsection{\bf IT phase}
The triangular lattice of NaVO$_{2}$ exhibits magnetic frustration and spatially
 anisotropic exchange interactions in the IT phase.
As shown in Table \Rmnum{2}, the \emph{G}-AF\Rmnum{1} configuration
 is the most stable state among the five magnetic structures both from
 LSDA and LSDA+U calculations. By a detailed analysis of the magnetic ground state \emph{G}-AF\Rmnum{1},
 see Fig. 2(a)(i), AFM chains are formed along the (\={1}10) direction (long V-V bonds),
 while AFM exchange is also more favorable
 along the (010) direction (short V-V bonds). Considering all the short V-V bonds
 are completely equivalent, both of the (100) and (010) directions should be AFM exchange.
 Thus, NaVO$_{2}$ in the IT phase can be regarded as a system with frustration effects.
  In addition, spatially
 anisotropic exchange interactions may exist in such a triangular spin lattice\cite{Giot}, i.e., \emph{J$_{1}$}
 along the direction of long V-V bonds and \emph{J$_{2}$}
along the two directions of short V-V bonds (Fig. 2(a)(i)).

 In order to describe the magnetic frustration and spatially
 anisotropic exchange interaction more clearly, we estimate the exchange
 interactions along one of the triangle directions (\emph{J$_{1}$}) and the other two (\emph{J$_{2}$}) (Fig. 2(a)(i)).
 Since all the configurations exhibit insulator characteristics, the spin
 size of V is stable, and the difference of total energy between
\emph{C}-AF\Rmnum{1} and \emph{G}-AF\Rmnum{1} (\emph{C}-AF\Rmnum{2}
and \emph{G}-AF\Rmnum{2}) configurations (See Table \Rmnum{2})
is so small that the system exhibits a 2D characteristic, a nearest neighbor Heisenberg-like Hamiltonian
may be a good primary approximation for the in-layer magnetic energy.
The corresponding 2D spin Hamiltommian can be written as
 \begin{equation}\label{(1)}
  H=J_{1}\sum_{(k,l)}\textbf{S$_{k}$}\cdot\textbf{S$_{l}$}+
  J_{2}\sum_{(i,j)}\textbf{S$_{i}$}\cdot\textbf{S$_{j}$}
 \end{equation}
where (\emph{i,j}) denotes a nearest-neighbor pair (short V-V bond) and (\emph{k,l}) denotes a
next-nearest-neighbor pair (long V-V bond). By mapping the obtained total
energies for each magnetic state to the Heisenberg model, the
exchange interactions \emph{J$_{1}$} and \emph{J$_{2}$} were calculated within this approximation:
\begin{equation}\label{2}
2\times(8\times4J_{2}S^{2})=E(FM)-E(\emph{C}-AF\Rmnum{2})
\end{equation}
\begin{equation}\label{3}
2\times(4\times4J_{2}S^{2}+4\times4J_{1}S^{2})=E(FM)-E(\emph{C}-AF\Rmnum{1})
\end{equation}
With S=1, we get \emph{J$_{2}$}=2.1 meV and \emph{J$_{1}$}=6.1 meV for NaVO$_{2}$
in the IT phase, which reflects strong spatial anisotropy. The AFM chains are
established along the (\={1}10) direction (\emph{J$_{1}$}) and the
interchain coupling (\emph{J$_{2}$}) is frustrated. Moreover, the
value of \emph{J$_{2}$}/\emph{J$_{1}$}=0.3 is so small that this magnetic structure
can be described as so-called weakly coupled \emph{zigzag} (S=1) chains
model\cite{Zheng}.

The integer spins (S=1) are able to weaken the frustration effects
in the frustrated systems, as in kagom\'{e} lattice\cite{Pati}.
Such a lattice has four nearest neighbors with the adjacent
triangles on the lattice sharing only one lattice point.
Interestingly, the triangular lattice can be composed of four kagom\'{e} lattices\cite{Kashibae}.
Thus, there are some analogous properties in these two frustrated systems.
Therefore, it is reasonable to suppose that the triangular lattice also has the
rule that the half-odd-integer spins are more highly frustrated than integer ones.
For example, NaTiO$_{2}$ ($S=\frac{1}{2}$)\cite{Ezhov}
and LiCrO$_{2}$ ($S=\frac{3}{2}$)\cite{Mazin} with half-odd-integer spins are always
frustrated even at low \emph{T}, while the magnetic frustration of NaMnO$_{2}$ (\emph{S}=2)
is clearly lifted by a structural distortion\cite{Giot}. Besides NaMnO$_{2}$,
NaVO$_{2}$ is another typical triangular lattice with integer spins (S=1).
Therefore, we can presume that the magnetic frustration in NaVO$_{2}$
can be lifted in some way.
\subsection{\bf LT phase}
From the discussion above, we expect that NaVO$_{2}$ with S=1 will
show a long-range magnetic ordering or a finite ground-state magnetization
 at low \emph{T}.
 As shown in Table \Rmnum{2}, \emph{G}-AF\Rmnum{2} is only 0.3 meV lower
 in total energy than \emph{C}-AF\Rmnum{2} within LSDA, and
both \emph{G}-AF\Rmnum{2} and \emph{C}-AF\Rmnum{2} have the same lowest total energy from LSDA+U
results for the LT phase. This reflects the obvious 2D
characteristic of NaVO$_{2}$: the interlayer interaction is much
weaker than the intralayer one. As stacking antiferromagnetically
between layers is observed in the experiment\cite{Mcqueen}, \emph{G}-AF\Rmnum{2} state should be more favorable at
low \emph{T}.
Such a magnetic state denotes the long range
3D magnetic ordering with AFM coupled FM chains in V cation layers and interlayer AFM coupling.
  Obviously, the magnetic frustration is lifted in \emph{G}-AF\Rmnum{2} state by
  the first-order transition at 91.5 K:  the lattice distorts to another monoclinic (\emph{C}2/\emph{m})
 with four long and two short V-V bonds reversed, comparing with IT phase.

The lattice distortion, which relieves the frustration, is driven by the
formation  of OO in the LT phase.
Fig. 4 shows the orbital characteristic of V 3\emph{d} in \emph{G}-AF\Rmnum{2} state.
Since the orbital occupancies of the two inequivalent V (V1 and V2 in Fig. 2(a)(ii))
 are nearly the same, only the density of states (DOS) of V1 3\emph{d} is shown.
 The \emph{z} axis of local coordinate system coincides with the V-O bond of the VO$_{6}$ octahedra.
 In such a coordinate system,
 \emph{d$_{zx}$} and \emph{d$_{yz}$} orbitals are mainly occupied and
\emph{ d$_{xy}$} orbital is less occupied at all V ions. Such an orbital
occupancy is consistent with the OO proposed by McQueen and Cava\cite{Mcqueen}:
the \emph{d$_{zx}$} and \emph{d$_{yz}$} orbitals are singly occupied with all unoccupied \emph{d$_{xy}$}
orbitals.

This OO relieves the magnetic frustration and stabilizes
the long-range magnetic ordering state.
In view of the weak superexchange interaction resulted from the nearly
\emph{ 90$^{\circ}$ } angle of V-O-V\cite{Goodenough} as well as the weak magnetic interaction between adjacent
 VO$_{2}$ planes interleaved by a layer of Na ions,
 the V-V direct exchange interaction in-plane should be dominant in such a particular crystal structure.
 Particularly, we only consider the $\sigma$ overlap in V-V direct exchange, which is much stronger than the
$\pi$ overlap. It means that each orbital in a V ion only hybridizes with the same orbitals
 in the two nearest-neighboring V ions. That is to say, \emph{d$_{yz}$} orbital hybridizes with two
 neighboring \emph{d$_{yz}$} orbitals in the (010) and (0\={1}0) directions,
 and \emph{d$_{zx}$} orbital hybridizes with two
 neighboring \emph{d$_{zx}$} orbitals in the (100) and (\={1}00) directions. The repulsions
 between the occupied orbitals (\emph{d$_{zx}$} or \emph{d$_{yz}$}) induce
 the elongation of V-V bonds along four directions: (010), (0\={1}0), (100) and (\={1}00).
 According to the Goodenough-Kanamori (GK) rules\cite{Good}, the strong AFM
 coupling should exist along these four directions because of the occupation of two orbitals with $\sigma$ overlap.
At the same time, the less occupancy of \emph{d$_{xy}$} orbital leads to
 V-V bonds contraction as well as a weak FM exchange along the (1\={1}0) and (\={1}10) directions. Thus,
the four long and two short V-V bonds result from bonding via \emph{d$_{zx}$}, \emph{d$_{yz}$}
orbitals, but no bonding of the \emph{d$_{xy}$} electrons. In other words,
such an OO results in the lattice distortion, and consequently relieves the magnetic frustration.

In the LT phase, another important aspect is that the SOC
 turns out to be crucial for the small magnetic
moment of 0.98 $\mu$$_{B}$ per V$^{3+}$ observed experimentally\cite{Mcqueen}. As shown in Table \Rmnum{2},
 the magnetic moments from LSDA+U calculations are
much larger than the ones observed in the experiment.
Further to investigate the magnetic moments changing with the
 particular choice of \emph{U$_{eff}$}, we calculate the moments
  for \emph{U$_{eff}$}=2-6 eV and find that the moments
  are not sensitive to \emph{U$_{eff}$}: as shown in Table \Rmnum{3},
  the magnetic moments stay constant within
 0.2 $\mu$$_{B}$ as long as the system is an insulator.
Since an easy magnetization direction (\={1}10) is observed
in the exprement\cite{Mcqueen}, the SOC may
play an important role in determining the total magnetic moment.
Thus, the SOC is included to reinvestigate the magnetic moment.

Then,
we perform LSDA+SOC+U calculations for the favorable magnetic configuration (\emph{G}-AF\Rmnum{2}),
 and obtain a local moment of 0.89 $\mu$$_{B}$ per V$^{3+}$
 with 1.65 $\mu$$_{B}$ spin and -0.77 $\mu$$_{B}$ orbital contributions.
 This value is half the expected moment (2 $\mu$$_{B}$), but very close to
 the observed one (0.98 $\mu$$_{B}$).
 The DOS projected on (2,m) space shown in Fig. 5 reveals the origin of orbital moment.
Note that the \emph{z} axis is set to the direction of easy magnetization along (\={1}10) now.
Since \emph{d$_{1}$} and \emph{d$_{-1}$} have nearly the same occupancies,
the orbital moment only comes from the contribution of
 different occupancies between \emph{d$_{2}$} and \emph{d$_{-2}$}.
  By further analysis, the \emph{d$_{2}$} occupancy is
 less than one half of the \emph{d$_{-2}$} one,
 which should give an orbital moment between 1 $\mu$$_{B}$ and 2 $\mu$$_{B}$.
 Nevertheless, there is no surprising that the calculated orbital moment is 0.77 $\mu$$_{B}$ here,
 because some hybridization effects are neglected in above analysis, e.g., the covalence effects
 with O 2\emph{p}. Thus, the inclusion of SOC leads to
 a surprising but experimentally sound results.

\section{\bf Conclusions }
In Summary, we have investigated the electronic structure and magnetic properties
 of NaVO$_{2}$ by first-principles calculations. The
 \emph{t$_{2g}$} orbitals are split into upper \emph{a$_{1g}$} and lower \emph{e'$_{g}$}
 states by a trigonal distortion of compression in the HT phase, which is similar to the splitting
 in Na$_{x}$CoO$_{2}$\cite{Landron2}.
In the IT phase, the crystal symmetry is lowered to \emph{C}2/\emph{m},
 under which the system behaves like a frustrated spin lattice
with spatially anisotropic exchange interactions.
Finally, a long-range ordered AFM ground state is formed when
 the magnetic frustration is relieved by another lattice distortion
  resulted from a certain ordering of occupied orbitals
at low \emph{T}.
 The small magnetic moment observed
 originates from the compensation of orbital moment for the spin moment.
 It is obvious that
 so many physical phenomena in the triangular lattice are reflected
 in NaVO$_{2}$, suggesting that NaVO$_{2}$ is a very good model material for studying
 2D triangular lattice systems.

\section{\bf Acknowledgements}
We thank T. M. McQueen, P. W. Stephens, Q. Huang, T. Klimczuk, F. Ronning, and R. J. Cava
for providing structural parameters prior to publication. This work was supported by
the special Funds for Major State Basic Research Project of China(973) under grant No. 2007CB925004, 863 Project,
Knowledge Innovation Program of Chinese Academy of Sciences, and  Director Grants of
CASHIPS, CUHK Direct Grant No. 2060345.
Part of the calculations were
performed in Center for Computational Science of CASHIPS and the Shanghai Supercomputer
Center.

\begin{table}[tb]
%============================== Table 1========================
\caption{The lattice parameters of NaVO$_{2}$ at 100, 91.5 and 20 K}

\begin{ruledtabular}
\begin{tabular}{cccc}

Temperature&100 K&91.5 K&20 K\\
 \hline
space group&\emph{R}-3\emph{m}&\emph{C}2/\emph{m}&\emph{C}2/\emph{m}\\

a({\AA})&2.9959(1)&5.1758(1)&5.2223(2)\\
b({\AA})&2.9959(1)&3.0078(1)&2.9755(1)\\
c({\AA})&16.0996(1)&5.6340(1)&5.6492(3)\\

$\alpha$($^{\circ}$)&90&90&90\\
$\beta$($^{\circ}$)&90&107.629(1)&108.335(1)\\
$\gamma$($^{\circ}$)&120&90&90\\

x,y,z(Na)&3a(0, 0, 0)&2a(0, 0, 0)&2a(0, 0, 0)\\
x,y,z(V)&3b(0, 0, 0.5)&2d(0, 0.5, 0.5)&2d(0, 0.5, 0.5)\\
x,y,z(O)&6c(0, 0, 0.2339(0))&4i(0.2368(7), 0, 0.6989(5))&4i(0.2296(5), 0, 0.7005(4))\\

\end{tabular}
\end{ruledtabular}
\end{table}
\clearpage
%==========================  end Table 1 ===========================

\begin{table}[tb]
%============================== Table 2========================
\caption{The total energy \emph{E} (meV/ (8f. u.)), magnetic moment \emph{M} ($\mu$$_{B}$)
per V$^{3+}$ and band gap \emph{E$_{g}$} (eV) in
different magnetic states.}

\begin{ruledtabular}
\begin{tabular}{ccccccccc}

&&Configuration&FM&\emph{C}-AF\Rmnum{1}&\emph{G}-AF\Rmnum{1}&\emph{C}-AF\Rmnum{2}&\emph{G}-AF\Rmnum{2}\\
 \hline
 &IT (T=91.5 K) &\emph{E}&514&4&0&128&132\\

LSDA&&\emph{M} &1.63&$\pm$1.37&$\pm$1.37&$\pm$1.34&$\pm$1.34\\
&LT (T=20 K) &\emph{E}&574&152&156&4&0\\
&&\emph{M} &1.51&$\pm$1.35&$\pm$1.35&$\pm$1.38&$\pm$1.38\\
\hline

&&\emph{E}&329&65&0&192&189\\
&IT (T=91.5 K) &\emph{M} &1.71&$\pm$1.65&$\pm$1.65&$\pm$1.66&$\pm$1.66\\
LSDA+U&&\emph{E$_{g}$} &1.1&1.2&1.1&1.4&1.5\\

&&\emph{E}&395&286&288&0&0\\
&LT (T=20 K) &\emph{M} &1.71&$\pm$1.66&$\pm$1.66&$\pm$1.65&$\pm$1.65\\
&&\emph{E$_{g}$} &1.2&1.4&1.4&1.1&1.1
\\
\end{tabular}
\end{ruledtabular}
\end{table}
\clearpage
%==========================  end Table 2 ===========================

\begin{table}[tb]
%============================== Table 3========================
\caption{The band gap \emph{E$_{g}$} (eV) and magnetic moment \emph{M} ($\mu$$_{B}$)
per V$^{3+}$ for different \emph{U$_{eff}$} (eV) in \emph{G}-AF\Rmnum{2} configuration.}

\begin{ruledtabular}
\begin{tabular}{cccccc}

\emph{U$_{eff}$}&2&3&3.6&5&6\\
 \hline
 E$_{g}$&0.02&0.9&1.1&1.8&2.3\\

\emph{M}&$\pm$1.57&$\pm$1.63&$\pm$1.65&$\pm$1.69&$\pm$1.71\\

\end{tabular}
\end{ruledtabular}
\end{table}
\clearpage
%==========================  end Table 3 ===========================

\begin{figure}[tp]
\vglue 1.0cm
\newpage

%---------------------------------Fig. 1-----------------

\caption {(color online). (a) The compressed octahedron of VO$_{2}$ layers. The
\emph{z} axis is the three fold axis of the VO$_{6}$ octahedron. $\alpha$
represents the O-V-O angle. (b) $\alpha$ angles in the HT, IT and LT phases.}

%-----------------------------------Fig. 2---------------

\caption {(a) Two different magnetic patterns in V-V plane
(i) \Rmnum{1}-type AF and (ii) \Rmnum{2}-type AF.
 The solid (dashed) lines
represent short V-V bonds in the IT (LT) phase and dashed (solid) lines
represent long V-V bonds in the IT (LT) phase.
\emph{J$_{1}$} (\emph{J$_{2}$}) denotes exchange interaction along the direction of dashed (solid) lines.
 (b) Schematic representation of
five magnetic configurations used in our
calculation. Only V atoms are drawn. Filled (open) circles indicate spin up (down) moments.}

%---------------------------------Fig. 3-----------------

\caption {The spin-majority band structure of NaVO$_{2}$
 from (a) LSDA and (b) LSDA+U (\emph{U$_{eff}$}=3.6 eV) calculations for the fixed structure at 100 K.
 The band with (a) 3\emph{d} and (b) \emph{a$_{1g}$} character is marked.}

%---------------------------------Fig. 4-----------------

\caption {Density of states (DOS) of NaVO$_{2}$
calculated by LSDA+U (\emph{U$_{eff}$}=3.6 eV) in the \emph{G}-AF\Rmnum{2}
state for the fixed structure at 20 K. Besides total 3\emph{d} state,
all the \emph{d$_{zx}$}, \emph{d$_{yz}$} and \emph{d$_{xy}$}
orbitals in local coordinate system for
V1 (Fig. 2(a)(i)) ion are depicted. Solid (dashed) lines denote the spin-up (down) states.}

%---------------------------------Fig. 5-----------------

\caption {Density of states (DOS) of NaVO$_{2}$
calculated by LSDA+U+SOC (\emph{U$_{eff}$}=3.6 eV) in the \emph{G}-AF\Rmnum{2}
state for the fixed structure at 20 K.
Solid (dashed) lines denote the spin-up (down) states.}

\end{figure}
\clearpage

%%%%%%%%%%%%%%%%%%%%%%The following is the figures%%%%%%%%%%%%%%%%%%%%%%%%%%%%%%%%%%%%%%%%%%%%%%%%%%%%%%%%%%%%
\newpage

\begin{figure*}[htbp]
\center {$\Huge\textbf{Fig. 1  \underline{tjia}.eps}$}

\includegraphics[]{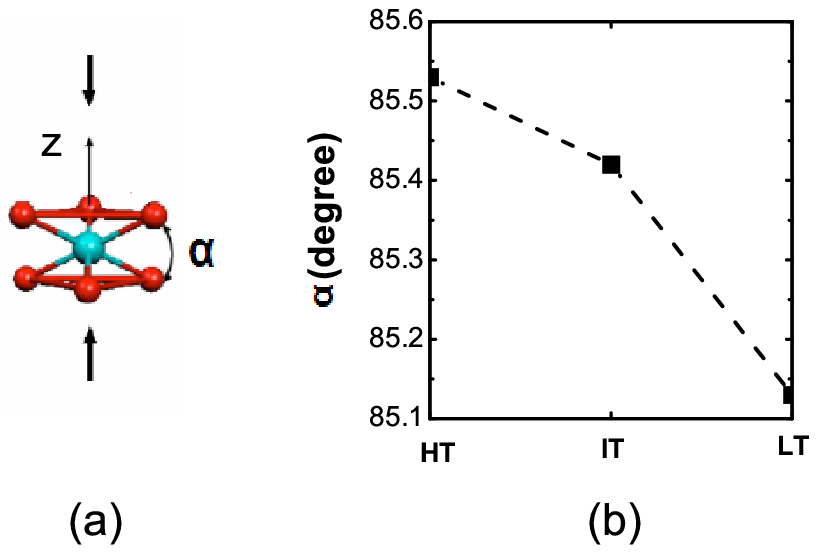}

\end{figure*}

\clearpage
\newpage

\begin{figure*}[htbp]

\center {$\Huge\textbf{Fig. 2  \underline{tjia}.eps}$}

\includegraphics[]{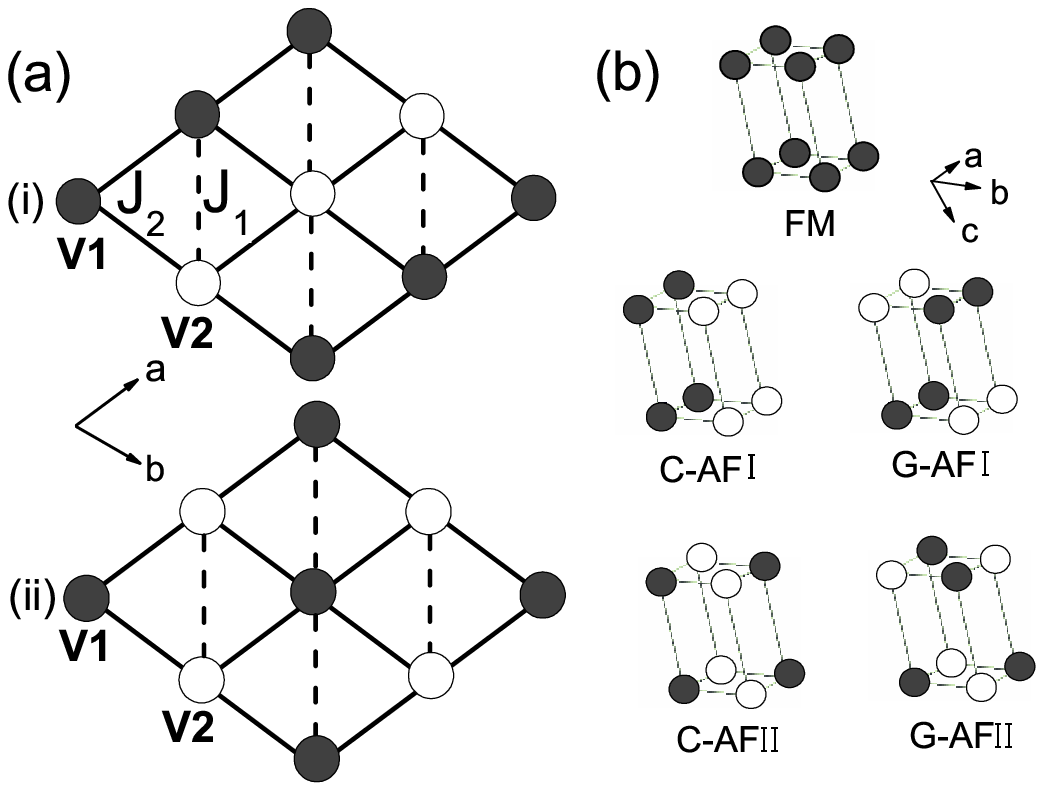}
\end{figure*}

\clearpage
\newpage

\begin{figure*}[htbp]
\center {$\Huge\textbf{Fig. 3  \underline{tjia}.eps}$}
\includegraphics[]{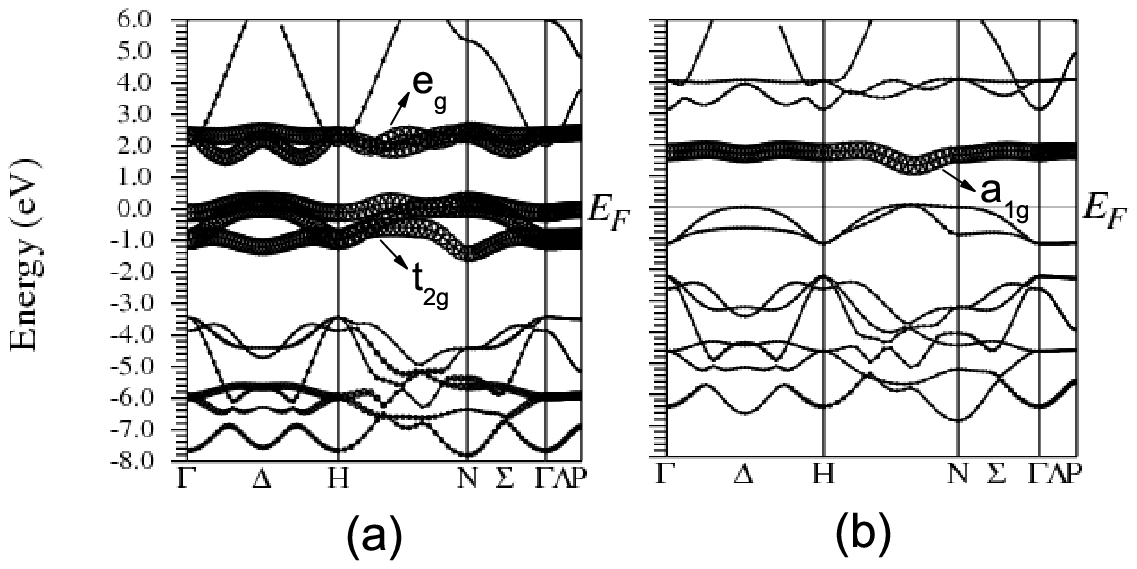}
\end{figure*}

\clearpage
\newpage

\begin{figure*}[htbp]
\center {$\Huge\textbf{Fig. 4  \underline{tjia}.eps}$}

\includegraphics[]{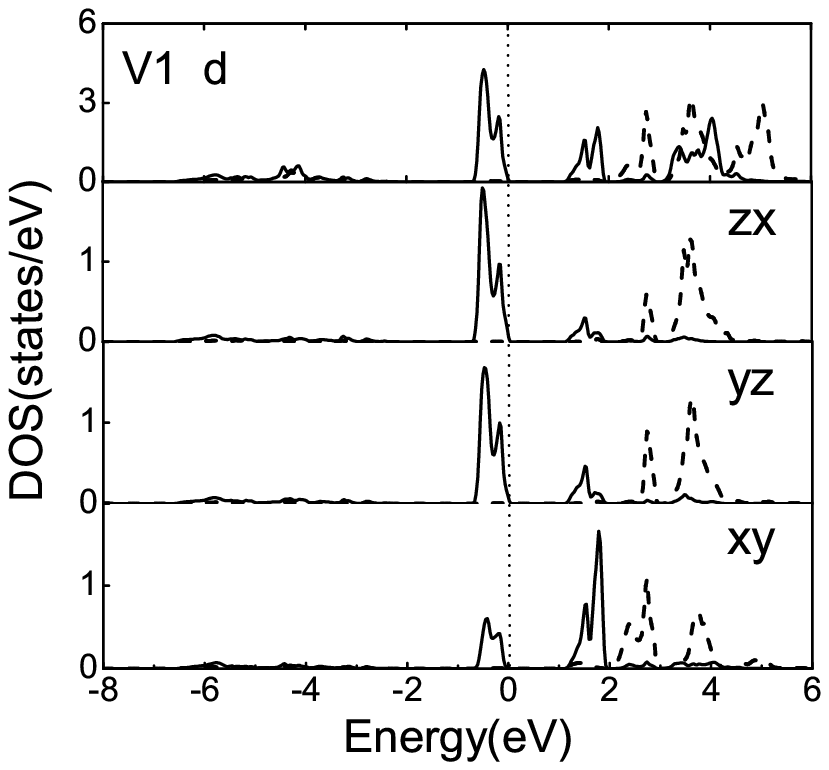}

\end{figure*}

\clearpage
\newpage

\begin{figure*}[htbp]
\center {$\Huge\textbf{Fig. 5 \underline{tjia}.eps}$}

\includegraphics[]{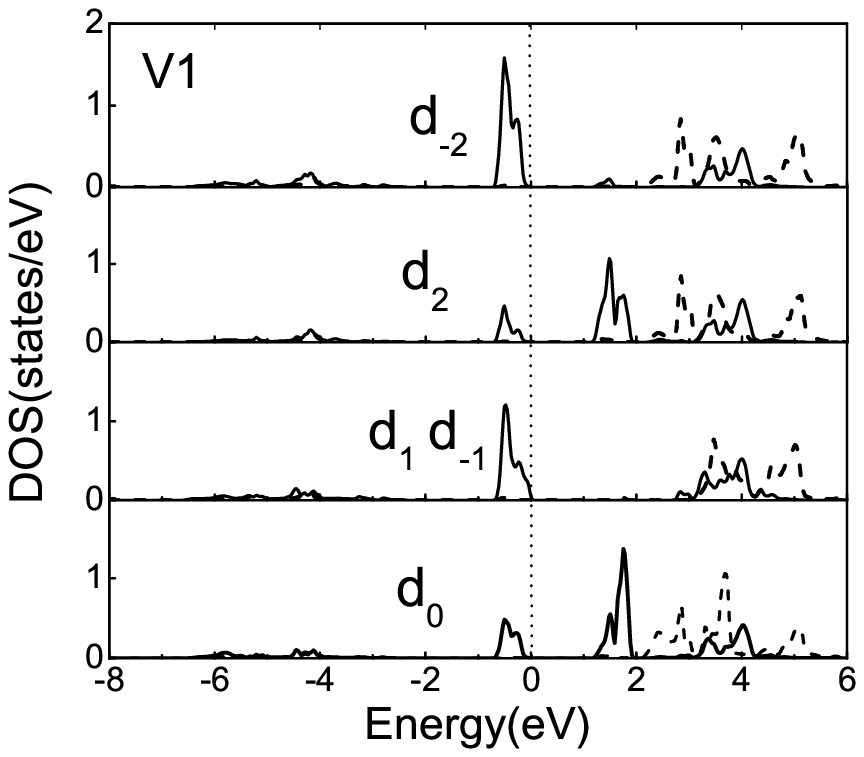}

\end{figure*}

\clearpage
\newpage


\begin{thebibliography}{00}
\bibitem{Takada}
K. Takada, H. Sakurai, E. Takayama-Muromachi, F. Izumi, R. A. Dilanian, and T. Sasali, Nature (London) {\bf422}, 53 (2003).

\bibitem{Giot}
Maud Giot, Laurent C. Chapon, John Androulakis, Mark A. Green, Paolo G. Radaelli, and Alexandros Lappas,
 Phys. Rev. Lett. {\bf99}, 247211 (2007).

\bibitem{Darie}
C. Darie, P. Bordet, S. de Brion, M. Holzapfel, O. Isnard, A. Lecchi, J. E. Lorenzo, and E. Suard, Eur. Phys. J. B
{\bf43}, 159 (2005).

\bibitem{Reynaud}
F. Reynaud, D. Mertz, F. Celestini, J.-M. Debierre, A. M. Ghorayeb, P. Simon, A. Stepanov,
J. Voiron, and C. Delmas, Phys. Rev. Lett. {\bf86}, 3638 (2001).

\bibitem{Ezhov}
S. Yu. Ezhov, V. I. Anisimov, H. F. Pen, D. I. Khomskii and G. A. Sawatzky,
Europhys. Lett. {\bf44}, 491 (1998).

\bibitem{Mazin}
I. I. Mazin, Phys. Rev. B {\bf75}, 094407 (2007).

\bibitem{Reitsma}
Albert J. W. Reitsma, Louis Felix Feiner and Andrzej M. Ole$\acute{s}$, New J. Phys. {\bf7}, 121 (2005).

\bibitem{Mila}
F. Mila, F. Vernay, A. Ralko, F. Becca, P. Fazekas and K. Penc, J. Phys.: Condens. Matter {\bf19}, 145201 (2007).

\bibitem{Chung}
J.-H. Chung, Th. Proffen, S. Shamoto, A. M. Ghorayeb, L. Croguennec, W. Tian, B. C. Sales, R. Jin, D. Mandrus
and T. Egami, Phys. Rev. B {\bf71}, 064410 (2005).

\bibitem{Tokura}
Y. Tokura and N. Nagaosa, science, {\bf288}, 462 (2000)

\bibitem{Yan}
J.-Q. Yan, J.-S. Zhou, and J. B. Goodenough, Phys. Rev. Lett. {\bf93}, 235901 (2004).

\bibitem{Horsch}
Peter Horsch, Andrzej M. Ole$\acute{s}$, Louis Felix Feiner, and Giniyat Khaliullin,
Phys. Rev. Lett. {\bf100}, 167205(2008).

\bibitem{Pen}
H. F. Pen, J. van den Brink, D. I. Khomskii, and G. A. Sawatzky
, Phys. Rev. Lett. {\bf78}, 1323 (1997).

\bibitem{Mcqueen}
T. M. McQueen, P. W. Stephens, Q. Huang, T. Klimczuk, F. Ronning, and R. J. Cava,
Phys. Rev. Lett. {\bf101}, 166402 (2008).


\bibitem{Onoda}
M. Onoda, J. Phys. Condens. Matter {\bf20}, 145205 (2008).

\bibitem{Pen2}
H. F. Pen, L. H. Tjeng, E. Pellegrin, F. M. F. de Groot, G. A. Sawatzky, M. A. van Veenendaal and C. T. Chen,
Phys. Rev. B {\bf55}, 15500 (1997).

\bibitem{interpretation}
The difference between 93 K and 91.5 K has to do with the sample being measured on warming or on cooling.
The IT to LT transition temperature is T=91.5 K here, corresponding to
the lattice parameters measured on cooling. And the slight lattice-parameter
differences from Ref. \cite{Mcqueen} are within experimental error.

\bibitem{Blaha}
P. Blaha\emph{ et al}., http: // www. wien2k. at.

\bibitem{Perdew}
John P. Perdew and Yue Wang,  Phys. Rev. B {\bf45}, 13244 (1992).

\bibitem{Anisimov}
Vladimir I. Anisimov, Jan Zaanen, and Ole K. Andersen, Phys. Rev. B {\bf44}, 943 (1991).

\bibitem{Dudarev}
S. L. Dudarev, G. A. Botton, S. Y. Savrasov, C. J. Humphreys and A. P. Sutton, Phys. Rev. B {\bf57}, 1505 (1998).

\bibitem{Madsen}
G. K. H. Madsen and P. Novak, Europhys. Lett. {\bf69}, 777 (2005).

\bibitem{Anisimov2}
V. I. Anisimov, I. V. Solovyev, M. A. Korotin, M. T. Czy$\dot{z}$yk and G. A. Sawatzky, Phys. Rev. B {\bf48}, 16929 (1993).

\bibitem{Laskowski}
Robert Laskowski, Peter Blaha, and Karlheinz Schwarz, Phys. Rev. B {\bf67}, 075102 (2003).

\bibitem{interpretation2}
We performed LSDA+U calculations for the two most stable
configurations (\emph{G}-AF\Rmnum{1} and \emph{G}-AF\Rmnum{2}) both in
the IT and LT phases with \emph{U$_{eff}$}=2, 3, 5, 6 eV.
The results show that the \emph{G}-AF\Rmnum{1} (\emph{G}-AF\Rmnum{2})
is always the most stable state in the IT (LT) phase
in such \emph{U$_{eff}$} range.

\bibitem{Landron}
Sylvain Landron and Marie-Bernadette Lepetit, Phys. Rev. B {\bf77}, 125106 (2008).

\bibitem{Kashibae}
W. Koshibae and S. Maekawa, Phys. Rev. Lett. {\bf91}, 257003 (2003).

\bibitem{Landron2}
Sylvain Landron and Marie-Bernadette Lepetit, Phys. Rev. B {\bf74}, 184507 (2006).

\bibitem{Zou}
Liang-Jian Zou, J.-L. Wang, and Z. Zeng, Phys. Rev. B {\bf69}, 132505 (2004).

\bibitem{Hasan}
M. Z. Hasan, Y.-D. Chuang, D. Qian, Y. W. Li, Y. Kong, A. P. Kuprin, A.V. Fedorov, R. Kimmerling,
E. Rotenberg, K. Rossnagel, Z. Hussain, H. Koh, N. S. Rogado, M. L. Foo, and R. J. Cava,
Phys. Rev. Lett. {\bf92}, 246402 (2004).

\bibitem{Zheng}
Zheng Weihong, Ross H. McKenzie and Rajiv R. P. Singh, Phys. Rev. B {\bf59}, 14367 (1999).

\bibitem{Pati}
Swapan K. Pati and C. N. R. Rao, J. Chem. Phys. {\bf123}, 234703 (2005).

\bibitem{Goodenough}
J. B. Goodenough, Phys. Rev. {\bf117}, 1442 (1960).

\bibitem{Good}
J. B. Goodenough, \emph{Magnetism and the Chemical Bond} (Interscience Publishers, New York, 1963);
J. Kanamori, J. Phys. Chem. Solids {\bf10}, 87 (1959).


\newpage
\end{thebibliography}
\end{document}